\documentstyle[aps]{revtex}

\input epsf

\begin{document}

\preprint{\vbox{\hspace*{\fill} DOE/ER/40762-087\\
          \hspace*{\fill} U. of MD PP \#96-112}} 


\title{Regularization, Renormalization and Range:
 The Nucleon-Nucleon Interaction from Effective Field Theory}

\author{Thomas D. Cohen}

\address{Department of 
Physics, University of Maryland, College Park, MD 20742-4111}

\maketitle

\begin{abstract}

Regularization and renormalization is discussed in the context of low-energy
effective field theory treatments of two or more heavy particles (such as
nucleons).  It is desirable to regulate the contact interactions from the outset by
treating them as having a finite range.  The low energy physical observables should
be insensitive to this range provided that the range is of a similar or greater
scale than that of the interaction. Alternative schemes, such as dimensional
regularization, lead to paradoxical conclusions such as the impossibility of
repulsive interactions for truly low energy effective theories where all of the
exchange particles are integrated out.  This difficulty arises  because a
nonrelativistic field theory with repulsive contact interactions is trivial in the
sense that the $S$ matrix is unity and the renormalized coupling constant  zero. 
Possible consequences of low energy attraction are also discussed.  It is argued
that in the case of large or small scattering lengths, the region of validity of
effective field theory expansion is much larger if the contact interactions are
given a finite range from the beginning.

\end{abstract} 

\section{Introduction}

One common issue in particle physics is the existence of phenomena on
widely differing energy scales.  In studying the low energy
phenomenology in such situations, the techniques of effective field
theory (EFT) have proven extremely useful\cite{EFT}.  They allow one
to include systematically only those effects of the short range
physics which contribute to the long range phenomena up to some given
level of accuracy.  The philosophy underlying this is that one can
integrate the short wavelength degrees of freedom, {\it i.e.} those
degrees of freedom whose momenta are larger than some separation
scale, $\mu$, out of the functional integral.  Of course, in doing
this one obtains an effective action which is nonlocal.  However, the
nonlocality is on the scale of the degrees of freedom which have been
integrated out.  At scales far below this it is legitimate to expand
this in the form of a derivative expansion.  It is often the case that
one cannot, in fact, carry out this partial functional integration of
the underlying fundamental theory either because it is technically
intractable or because one does not know the underlying theory in
detail.  In this case, one can use a knowledge of the form of the
symmetries of the underlying theory to develop an effective field
theory with phenomenological coefficients which corresponds to
the derivative expansion of the full theory.  A classic example of
this approach is chiral perturbation theory which has been used to
describe the interactions of pseudo-Goldstone bosons with each other
\cite{GL}.

Several years ago, Weinberg suggested that the technology of EFT---when
properly modified---could be used to describe low energy
nuclear phenomena such as nucleon-nucleon scattering and bound states and
the interaction of nuclei with pions and photons\cite{Weinberg}. 
The key to this approach was 
the development of a formalism based on a systematic power counting scheme
describing the interactions  of heavy particles ( where 
``heavy'' means that the
mass is very large compared to the momentum scale being probed).  The
fundamental insight is that the power counting should apply to n-particle
irreducible graphs ({\it i.e.} potentials) and not to the full  amplitudes.
The full amplitudes are obtained by iteration of these  potentials.
The approach is implemented via an effective Lagrangian containing explicit
light degrees of freedom ({\it e.g.} pions) along with contact interactions
whose coupling constants serve to parameterize the effects of shorter range
physics.  Weinberg's suggestion has 
inspired a considerable amount of research on effective field theoretic
approaches to low energy nuclear
phenomena\cite{w1,w2,w3,w4,w5,w6,w7,w8,w9,w10,w11,w12,ksw}.

In this paper it will be shown that great care must be exercised when
renormalizing this effective theory. A version of the formalism elucidated by
Weinberg has a rather perverse feature which can be traced to the
renormalization scheme: the approach is apparently incapable of
describing systems whose low energy interactions are repulsive in the
limit of very low energy scattering; {\it i.e.}, the limit where the
momenta are much less than all of the masses in the problem (so that in
the nuclear case one could integrate out the  pion).  In such a
case, as discussed in refs. \cite{Weinberg} and \cite{ksw} one can
integrate out all of the light degrees of freedom to obtain an
effective lagrangian with contact interactions only.  To lowest order
in the power counting, the $T$ matrix for s-wave scattering of heavy
fermions ({\it e.g.} nucleons) in Weinberg's treatment\cite{Weinberg}
depends on only a single parameter which corresponds to a particular
combination of spin-independent and spin-dependent contact
interactions whose renormalized value is fixed by the scattering
length, $a$: 
\begin{equation} T_0 (p^\prime, p) = \frac{4 \pi/M}{1/a +
i \sqrt{M E \, + \, i \epsilon}} \; \; \; \; ,
\label{T} 
\end{equation} 
where $M$ is
the mass of the particles and $p$ is the magnitude of the momentum of
the nucleon in the center of mass frame.  The subscript, 0, indicates
that this $T$ matrix was derived from the contact interaction with no
derivatives.  The energy of the state, $E$, is $p^2/M$ for scattering
states and the $i \epsilon$ fixes the boundary conditions in
extrapolations to negative energies.

The difficulty is easy to see from  eq.~(\ref{T}).
Elementary considerations show that a negative value of $a$
necessarily corresponds to  attraction.  On the other hand, a
positive value of $a$ can either correspond to repulsion or to
attraction with at least one bound state.  Bound states give rise to
poles in the $T$ matrix for negative energies.  Purely repulsive
interactions always correspond to a $T$ matrix without negative
energy poles.  From the form of the $T$ matrix in eq.~(\ref{T}),
however, it is apparent that when $a$ is positive, there is always a
pole in the $T$ matrix at $E= - 1/(M a^2)$.  Thus, regardless of the
sign of $a$, the $T$ matrix in eq.~(\ref{T}) corresponds to an
attractive interaction.

How serious a problem is this? One might argue that the problem is
purely formal and is of no phenomenological concern. After all, in
nuclear physics the potential {\it is} attractive at low energies; the
inability to describe repulsion may simply not be relevant.  On the
other hand, the EFT methods used to derive eq.~(\ref{T}) are not
particular to nuclear physics and never explicitly use the fact the
interaction is attractive---if the arguments are valid they ought to
apply equally well to cases where the interaction is repulsive.
Nothing in Weinberg's power counting scheme depends on the sign of the
interaction. Thus, the inability to describe repulsion suggests that
something is seriously wrong with the formalism.  As will be seen in
this paper the difficulty can ultimately be traced to the fact that the
interaction in the effective Lagrangian has zero range.  The only way
which an explicit range can enter into the dynamics in this approach
is through regulation and renormalization prescriptions.  The general
issues of regulation and renormalization are clearly important in the
attractive case.

It will be shown here that the problem is technical and is related to
the renormalization scheme used in the derivation of eq.~(\ref{T}).
It should be recalled that the contact terms in an effective
Lagrangian do not, in fact, describe zero range physics.  Rather, they
serve to parameterize the effects of physics of shorter range than the
separation scale.  Ultimately, the contact terms lead to divergences
which necessitate some regularization prescription and  an
associated renormalization of the coefficients in the lagrangian.
The regularization prescription should be consistent with the fact
that the interactions are, in fact, of finite range.  For example, one
can introduce a regulator into the contact interaction, thus making it 
a finite range interaction.  The range of this interaction should not
be taken to be zero in any intermediate step of the calculation. At
the end of the calculation, the regulator parameter should be fixed by the
separation scale, $\mu$.  As will be discussed here, it must
correspond to a larger range than the typical range of the potential
({\it e.g.} the effective range).  If there is a true separation of
scales in the problem, one will find that low energy physical
observables will be insensitive to the precise choice of the
separation scale and the form of the regulator.

In the derivation of Eq.~(\ref{T}), however, it was implicitly assumed
that the range is, in fact, zero.  That is, at various points in the
calculation the contact interaction is treated literally, as opposed
to merely serving to parameterize some short range physics.  It has
been known for some time that the repulsive $\delta$ function
interaction in nonrelativistic quantum mechanics is trivial---the
renormalized coupling constant must be zero and the $S$ matrix,
unity\cite{trivial}; this is a consequence of Friedman's theorem
\cite{Friedman}. Thus, it is not surprising that Eq.~(\ref{T}) fails
to describe repulsion.  Treating the contact terms literally is
incompatable with the derivation of the effective field theories from an
underlying theory since integrating out short range physics yields a
nonlocal theory.  Of course, in most applications of EFT this
inconsistency is innocuous in that errors induced by it are small and
can be systematically corrected at higher orders.  However, in the
case of two heavy particles where one must iterate the potential to
all orders the problem can be serious.  The inability to
describe repulsion should be viewed as an artifact of this inconsistent
treatment.

It is important to use a consistent regularization scheme even in the
case of attractive interactions.  For example, as recently noted in
Ref. \cite{ksw}, the convergence of Weinberg's scheme is controlled
by the scattering length; as the the scattering length diverges the
region of validity of the expansion tends to zero.  In nature the
scattering length is quite large, implying a very limited regime of
applicability of the approach.  As will be discussed briefly in this
paper and in more detail in a subsequent work this is also a
consequence of a renormalization scheme based on truly zero-ranged
interactions.  The central point of this paper is that if one wishes
to use EFT methods in nuclear interactions it is essential to regulate
the contact interactions from the outset by giving them finite range.

It is worth noting that excepting work based on a new expansion scheme
proposed in Ref. \cite{ksw}, numerical studies of the $NN$ force based
on effective field theories and chiral counting do not employ the
renormalization prescription used in the derivation of Eq.~(\ref{T}).
Rather, they cut off the integrals in the momentum-space Schr\"{o}dinger equation 
which effectively gives a finite range to the interactions.  Thus the
problems discussed here do not afflict the calculations in Refs.
\cite{w1,w4,w10}

\section{The Low Energy  $T$ Matrix} \label{DT}

Before discussing the problem of repulsion in any detail it is useful
to review how the $T$ matrix in Eq.~(\ref{T}) emerges in an effective
field theory treatment.  In order to use effective field theories one
needs a systematic power-counting scheme.  Traditionally in effective
field theory treatments this power counting is for a Feynman
amplitude.  However, as pointed out in ref \cite{Weinberg}, such a scheme
fails for the situation where two or more heavy particles interact
strongly at low energy.  The difficulty is that if the particles
typically have a momentum $Q$, the free propagator goes as $M/Q^2$ and
becomes large in the limit of small $Q$ destroying simple power counting
in $Q/\Lambda$.  The solution to this is quite simple---instead of
using power counting for the Feynman amplitude itself one develops a
systematic power counting only for the n-particle irreducible
graphs---{\it i.e.} for potentials.  The details of the power counting
argument will not be given here as it is well described in Ref.
\cite{Weinberg}.

To obtain scattering amplitudes, one can iterate these potentials to
all orders which corresponds to solving the Schr\"{o}dinger equation for
these potentials.  Recently Kaplan, Savage and Wise (KSW) have
proposed a different resummation\cite{ksw}, in which the lowest order
potential is summed to all orders as a Schr\"{o}einger equation and
subsequently the inverse of the real part of the Feynman amplitude is
expanded systematically.  This apparently greatly improves the
convergence of the expansion when the scattering length is large.
However, the problem discussed here applies to the lowest order
calculation of the scattering amplitude and it affects both the
Weinberg and the KSW schemes.

In the case where all of the particles are treated as heavy, including all
exchanged bosons, it is trivial to write down the potentials
to some order. They are given  in terms of an effective Lagrangian 
which consists entirely of contact interactions with various numbers
of derivatives. This effective Lagrangian is:
\begin{equation}
 {\cal L} = N^\dagger \,  i \partial_t \, N \, -  \, N^\dagger \, 
 \frac{\nabla^2}{2M} \, N \,
 - \, \frac{1}{2} \,  C_S (N^\dagger N)^2  - \, 
 \frac{1}{2} \,  C_T (N^\dagger \, \vec{\sigma} \, N)^2 \ldots \; \; \; \; ,
\end{equation}
where $\ldots$ indicates contact terms with two or more derivatives.
Such terms are higher order in the power counting.  Isoscalar s-wave
scattering only depends on the combination $C \equiv (C_S - 3 C_T)$.

The next step is to solve the Schr\"{o}dinger equation with appropriate
boundary conditions for scattering and thus  determine the $T$
matrix.  This is done most naturally in the form of the Lippmann-Schwinger
equation: $T = V + V G_0 T$, where $G_0 = 1/(E - p^2/M + i
\epsilon)$ and $p$ is the relative momentum operator.  Clearly, this
corresponds to iterating the potential to all orders.  As written
above, the Lippmann-Schwinger equation is an operator equation; in
momentum space it is an integral equation:
\begin{equation}
T(p,p^\prime) \, = \, V(p,p^\prime) + (2 \pi)^{-3} \int \, {\rm d}^3
p^{\prime \prime}
V(p, p^{\prime \prime}) \,  G_0(p^{\prime \prime}; E) \, T(p^{\prime
\prime},p^\prime) \; \; \; \; ,
\label{LS}\end{equation}
where $ G_0(p; E) = 1/(E - p^2/M + i \epsilon)$.  For an arbitrary $V$
one must solve this equation via standard numerical means.  

For the present case the zeroth order potential is simply a delta
function in configuration space and therefore a constant in momentum space; 
$V_0(k,k^\prime) = C$.  Formally, it is straightforward to solve the
Lippmann-Schwinger equation with this potential.  Since $V_0$ is a
constant the equation becomes algebraic; the solution is
\begin{equation} T_0(p,p^\prime) \, = \, \frac{1}{1/C - (2 \pi)^{-3}
\int \, {\rm d}^3 p^{\prime \prime} G_0(p^{\prime \prime},E)} \; \; \; \; .
\label{formsol}
\end{equation}
Unfortunately, the solution is only
formal since $(2 \pi)^{-3} \int\, {\rm d}^3 p^{\prime \prime}
G_0(p^{\prime \prime},E)$ diverges, and so as written the solution is
meaningless. This is hardly surprising--- it is well known that in 3+1
dimensions, delta function potentials with finite strength
 are sufficiently singular as to
have no well-behaved solutions.

Thus, to make sense of eq.~(\ref{formsol}) one must renormalize.  The
bare parameter 
$C$ must go to zero, but must do so in such a way that the $T$ matrix
remains finite. 
Weinberg introduces a renormalized coupling $C_R$ given by
\begin{equation} 
1/C_R \, = \, 1/C \, - \, (2 \pi)^{-3} \, \int {\rm d}^3 p^{\prime \prime} 
G_0(p^{\prime \prime}, E=0) \; \; \; \; .
\label{RN} \end{equation}
In terms of $C_R$ the $T$ matrix is given by
\begin{equation}
T_0(p,p^\prime) \, = \, \frac{1}{1/C_R - (2 \pi)^{-3} \int\, {\rm d}^3
p^{\prime \prime}
\left [ G_0(p^{\prime \prime},E)-G_0(p^{\prime \prime},E=0) \right ]} \, = \,
 \frac{4 \pi}{4 \pi /C_R + i M \sqrt{M E + i \epsilon}} \; \; \; \; .
\label{final} \end{equation}
The second equality is easily obtained since the integral is now convergent.
Finally, identifying the zero energy $T$ matrix as $4 \pi a/M$ immediately
gives a renormalization condition that $C_R = 4 \pi a/M$ and  yields
Eq.~(\ref{T}).
It is also worth observing at this stage
Weinberg's renormalization scheme is completely equivalent to
dimensional regularization with the $\overline{\rm MS}$ renormalization 
scheme as discussed in KSW.
  
\section{Regularization, Renormalization, Range and Repulsion}

This section addresses the question of why the calculation based on the
renormalization prescription discussed in Sec. {\ref{DT}} cannot
 describe repulsion.  As mentioned in the Introduction,
this occurs because the calculation implicitly assumes that the range of
the interaction is zero and not simply shorter than some separation
scale.  One indication Eq.~(\ref{T}) is based on a true zero range
interaction is the absence of any dependence on a regulator mass in
the final expression for the $T$ matrix.  Indeed, in Weinberg's
derivation no regularization scheme is explicitly introduced.  In
fact, the regulator mass has implicitly been taken to infinity at two
distinct places in this calculation.  The first is the derivation of
Eq.~(\ref{formsol}); had a finite range been given to the interaction
via any form of a regulator, one could not obtain the simple result of
Eq.~(\ref{formsol}).  Instead one would have had to solve an integral
equation.  The second place where the regulator mass was implicitly
taken to infinity is in the second equality in Eq.~(\ref{final}).

KSW reproduce Weinberg's result using dimensional regularization.
This, is to be expected, since by construction, dimensional
regularization introduces no regulator mass. In principle, a scale can
enter the problem through renormalization but, as noted by KSW, at this
order the renormalization scale dependence is trivial:
\begin{equation}
\mu \, \partial_\mu \, (1/C_R) \, = \, 0 \; \; \; \; .
\end{equation}
The KSW result is the same  as Weinberg's and suggests that the
lack of a regulator in the derivation of Eq.~(\ref{formsol}) is
sufficient for the system to lose the  information that the range of the
interaction  is finite. 

To see that Eq.~(\ref{T}) does correspond to a truly zero-range
interaction one should study finite ranged interactions and then show
that Eq.~(\ref{T}) is the zero-range limit.  Consider a regularization
prescription where one replaces the $\delta$ function potential by a
finite-ranged potential at the beginning of the problem.  If one is in
the regime in which the effective field theory is valid, then the
results are insensitive to the precise form of the regulator and the
precise value of the regulator mass.

  For simplicity, consider a simple form for the
regulated $\delta$ function---a square well of radius   $1/\mu$: 
\begin{equation}
\delta_R(\vec{x}; \mu) \, = \, \frac{3 \, \mu^3 \, 
\theta ( 1/\mu \, - \, |x|)}{4 \,\pi} \; \; \; \; ,
\label{delreg}
\end{equation}
where $\mu$ is the regulator mass. In coordinate space, the potential is just 
\begin{equation}
 V_0(\vec{x}) \, = \, C(\mu ) \, \delta_R(\vec{x}; \mu) \; \; \; \; .
\end{equation}
The bare coefficient is written as $C(\mu)$
to indicate that the value of the coupling depends on the regulator mass,
$\mu$, through a renormalization condition.

It is an elementary exercise to find the $T$ matrix associated with
this potential. The phase shifts satisfy 
\begin{equation}
p \, \cot (\delta) \, = \, \frac{ \kappa \, \cot (\kappa/\mu) \, + \, p \,
\tan(p/\mu)}{1 \, - \, \kappa/p \, \cot(\kappa/\mu) \, \tan(p/\mu)} \; \; \; \; ,
\label{ps}
\end{equation}
with
\begin{equation}
\kappa = \sqrt{p^2 - \frac{3  \, C(\mu) \, M \mu^3}{4 \pi}} \; \; \; \; .
\label{kap}
\end{equation}
This expression is valid for both attractive and repulsive interactions.  For
repulsive interactions and sufficiently small $p$, $\kappa$ becomes imaginary.
The on shell $T$ matrix is related to $\cot (\delta)$ by 
\begin{equation}
T(p) \, = \, = \, \frac{-2 \, \pi}{M \, p \,  ( \cot (\delta)
\, - i )} \; \; \; \; .
\label{tgeneral} 
\end{equation}

The expression for the phase shift in Eqs.~(\ref{ps}) and (\ref{kap})
depends on 
the bare coupling $C(\mu)$.  It is useful to express this in terms of a
physical observable.  This amounts to picking a renormalization condition
for $C(\mu)$.  The most natural choice is to use the scattering length 
which is related to the phase shifts near $p=0$:
\begin{equation}
\lim_{p \rightarrow 0} p \, \cot (\delta) \, = \, -1/a
\label{sl}
\end{equation} 
to fix $C(\mu)$.  Using Eqs.~(\ref{ps}), (\ref{kap}) and (\ref{sl}), one finds
the following renormalization condition for $C(\mu)$:
\begin{equation}
\sqrt{ - \frac{3 \, C(\mu ) \, M \,  \mu}{4 \pi}} \, \, 
\cot \left( \sqrt{ - \frac{3 \, C(\mu ) \, M \, \, \mu}{4 \pi}} \right) \,= 
 \,\frac{1}{1 \, - \,  a \,\mu} \; \; \; \; .
\label{Crenorm}
\end{equation}

It is straightforward to  demonstrate that for attractive interactions
in the limit of $\mu \rightarrow \infty$, one recovers Eq.~(\ref{T}). 
The key point is that in this limit $-C(\mu) \, \mu^3 \rightarrow \infty$
and thus $\kappa$ also diverges.  Moreover, as $-C(\mu ) \mu^3
\rightarrow \infty$,  $\kappa \rightarrow  ( - \frac{3 \, C(\mu ) \,
\mu^3}{4 \pi})^{1/2}$  which is  independent of $p$. Although $C(\mu ) \mu^3$
diverges, $C(\mu ) \mu$ can remain finite.
 Moreover $p/\mu \rightarrow 0$.   Imposing the
limit, one finds  
that Eq.~(\ref{ps}) becomes
\begin{equation}
\lim_{\mu \rightarrow \infty \, , \, C(\mu)\mu \, {\rm fixed}}
 \, p \, \cot (\delta) \, = \, 
 \frac{ ( - \frac{3 \, C(\mu ) M \, \, \mu^3}{4 \pi})^{1/2} \, 
\cot [( - \frac{3 \, C(\mu ) \, M \, \mu}{4 \pi})^{1/2}] }{1 \, - \,
 ( - \frac{3 \, 
C(\mu ) \, M \,  \mu}{4 \pi})^{1/2}  \,
 \cot[( - \frac{3 \, C(\mu ) \, M \,  \mu}{4 \pi})^{1/2}]}
\label{pcot}
\end{equation}
where the right-hand side of Eq.~(\ref{pcot}) is independent of $p$.  Imposing 
the renormalization condition in eq.~(\ref{Crenorm}) on the expression in
Eq.~(\ref{pcot})  one sees that $p \cot (\delta) = -1/a$;
Eq.~(\ref{T}) immediately follows.  The conclusion of this analysis is
that, as expected, Eq.~(\ref{T}) corresponds to an interaction of
literally zero range. 
 
Now consider what happens for a repulsive potential with $C(\mu) >0$.
Formally, Eq.~(\ref{pcot}) still applies.   There is a difficulty,
however, in implementing the renormalization  condition.  For $C(\mu)
> 0$, Eq.~(\ref{Crenorm}) becomes 
\begin{equation}
\sqrt{ \frac{3 \, C(\mu ) \, M \,  \mu}{4 \pi} } \, \,
\coth \left ( \sqrt{ \frac{3 \, C(\mu ) \, M \,\mu}{4 \pi}} \right ) \, 
 = \, \frac{1}{1 \, - \, a \, \mu} \; \; \; \; .
\label{repulrenorm} 
\end{equation}
For repulsivive interactions, $C(\mu) > 0$ and the left-hand side of 
Eq.~(\ref{repulrenorm}) is  positive
 so that  the renormalization condition can only be
satisfied  if 
\begin{equation}
 \mu < 1/a \; \; \; \; .
\end{equation}
Thus, when describing repulsion, one cannot take  the regulator mass 
to infinity while still describing the correct scattering length.
Indeed, when one lets $\mu \rightarrow \infty$ one is forced to have
$a \rightarrow 0$ which implies a zero cross section; as $\mu \rightarrow
0$, all effects of the repulsive interaction must vanish.  
 
Of course, the preceding analysis is just an alternative demonstration of the
triviality of  the repulsive delta function interaction discussed in
the context of 
the nonrelativistic limit of $\phi^4$ field theories by B\'eg and
Furlong \cite{trivial}.  A rigorous mathematical proof of this was
provided by Friedman \cite{Friedman}.
 
There is no great mystery here.  A regulated delta function of the
form in Eq.~(\ref{delreg}), with an infinite strength repulsive
interaction is simply a hard core interaction of radius $1/\mu$.  The
scattering length for a hard core potential is just the radius of the hard
core.  Thus, no matter how strong the repulsion in the regulated
$\delta$ function, one cannot get a scattering length greater than
$1/\mu$.  It is very clear why this happens, as $\mu \rightarrow
\infty$, $C(\mu)$ gets large.  The effect of a potential which has a
large positive value over some finite region is simply to exclude the
wave function from that region.  As $\mu \rightarrow \infty$, however,
the size of the region over which the wave function is excluded goes to
zero and the effect of the repulsion vanishes.
 
It is worth stressing that Friedman's theorem guarantees that the
inability to describe repulsion when one takes the regulator mass to
infinity is a general feature and not simply a peculiar feature of the
square-well regulator.  This can be explicitly verified by choosing
various alternative forms.  For example, the regulated delta function
can be chosen to be a surface delta function on a shell of radius
$1/\mu$: 
\begin{equation}
 \delta_R(\vec{x}; \mu) \, = \,
\frac{\mu^2}{4 \pi} \delta (|x| \, - \, 1/\mu) \; \; \; \; .
\label{sdi}
\end{equation} 
Taking $V_0(\vec{x}) \, = \, C(\mu) \,
\delta_R(\vec{x}; \mu)$, calculating the $T$ matrix and using the
scattering length to fix $C(\mu)$ gives the following renormalization
condition: 
\begin{equation} 
C(\mu) = \frac{4 \, \pi \, a}{M \, (\, 1
\, - \, \mu a )}  \; \; \; \; .
\end{equation} 
As in the case of the square well
regulator, one can  satisfy the renormalization condition for
repulsive interactions (which of necessity have $C(\mu) >0$ and $a
>0$) only for $\mu < 1/a$.

\section{Attractive Interactions and The Convergence Of the EFT Expansion }

The preceding section showed that, in order to describe repulsion in an
effective field theory with all exchanged particles integrated out, it
was necessary to regulate the theory by giving the contact
interactions a finite range.  Moreover, it was seen that it was not
possible to let the regulator parameter  go to infinity.  
This section briefly discusses
 possible consequences of taking the regulator mass to infinity for
attractive interactions.  It is easy to see that the problems
arise with such a scheme when the scattering length is either very
large or very small.  The case of large scattering length is of
particular importance since in the nuclear physics case the scattering
length in the singlet channel is very large.  This situation was 
discussed by KSW who point out that
Weinberg's scheme, when implemented with dimensional regularization
and $\overline{\rm MS}$ renormalization, breaks down at a momentum
scale set by the  scattering length.  As the scattering length goes
to infinity, Weinberg's approach breaks down for lower and lower
momentum; if $a$ were infinite Weinberg's expansion would break down
for arbitrarily small $p$ and thus be of no utility.

KSW suggest that this breakdown is a consequence of strong
correlations between the coefficients of contact terms at different
orders in the EFT expansion of the potential.  They propose to avoid
this difficulty by expanding $p \cot{\delta}$ rather than by expanding the
potentials and iterating to all orders as proposed by Weinberg.  At
first glance the explanation for the breakdown of Weinberg's scheme
seems quite unnatural; it depends on a conspiracy among the higher
order terms.  On the other hand, one might argue that generically
the scattering length should be of order $1/\Lambda$ and that having a
very long scattering length---one much longer than $1/\Lambda$---is,
in itself, unnatural.  Thus, one might expect that to describe such a
situation an {\it a priori} unlikely correlation among various terms
in the expansion is not absurd.  However, even if there are
correlations of the form postulated by KSW, there is still a problem.
The conventional power counting scheme requires  that the
contribution of $V_2$, the two derivative contact interaction, to the
$T$ matrix  be down by a power of $p^2/\Lambda^2$, compared to the
effect of $V_0$; this should hold up to momenta of order $\Lambda$.
KSW show explicitly that this fails for large $a$ when dimensional 
regularization and $\overline{\rm MS}$ renormalization is used.  This
raises a thorny question since there is no obvious flaw with conventional
power counting arguments and the power counting does not obviously depend
on the scattering length being small.

In this section, an alternative explanation for the breakdown of
Weinberg's scheme at low $p$ for large $a$ will be explored. It will
be argued that the breakdown is another consequence of taking the 
regulator mass to infinity and is not an intrinsic defect in the expansion.

In many ways, this problem is quite analogous to the difficulty of
describing repulsion.  In the repulsion case, the range of the
interaction was intrinsic to the description---the scattering length
was always smaller than the range of the potential.  Thus any scheme
which treats the range as being zero is destined to fail.  The
effective range in the case of infinite scattering length is similar. 
Recall that the effective range,  $r_0$, is defined in terms of an expansion of
$p \cot (\delta)$,
\begin{equation}
p \,\cot (\delta) \, = \, -1/a \, + \, 
\frac{1}{2} \, r_0 \, p^2 \, + \,\ldots  \; \; \; \; .
\label{r0}\end{equation}
 Suppose for example that the underlying dynamics were in fact a
square well.  Then it is trivial to show from Eqs.~(\ref{ps}), (\ref{kap}) 
and (\ref{r0})
that when the scattering length is infinite,  the effective range is
just the radius of the well. Thus, the physical size of the well is an essential
part of the physics of the effective range when $a$ is infinite.  It
will hardly be surprising if it turns out not to be possible to
describe this by a zero-range interaction.

Consider the treatment of the physics of the effective range in
Weinberg's scheme.  Clearly it depends on $V_2$, the two-derivative
contact term in the effective Lagrangian.  Formally, the effects of
this are order $p^2/\Lambda^2$ suppressed relative to $V_0$.  Although
there are several terms in the Lagrangian of this order, only one
linear combination plays a role in the singlet s-wave channel and one
can write $V_2$ as
\begin{equation}
V_2(\vec{p^\prime},\vec{p}) \, = \, \frac{C_2}{2} (p^2 + p^{\prime
\,2}) \; \; \; \; .
\end{equation}
Iterating this potential, using dimensional regularization and
$\overline{\rm MS}$ 
renormalization and using the scattering length $r_0$ to fix the
renormalized $C_2$ 
gives the following $T$ matrix \cite{ksw}:
\begin{equation}
T_2(p^\prime, p)  =  \frac{4 \pi/M}{\left (a + \frac{1}{2} a^2 r_0 p^2
\right )^{-1} 
\,+ \, i \sqrt{M E + i \epsilon}}\; \; \; \; .
\label{T2}
\end{equation}
The subscript, 2, indicates that this T matrix includes the effects of contact
interactions with up to two derivatives.

By conventional power counting one expects $T_2 \, = \, T_0 \, [ 1 \,
+ \,{\cal O}(p^2/\Lambda^2)]$.  However, expanding Eqs.~(\ref{T2})
and comparing with eq.~(\ref{T}) one sees that
\begin{equation}
T_2(p^\prime, p) \, = \, T_0(p^\prime, p) \, [ 1 \, + \, \frac{1}{2} a
r_0 p^2 \, + {\cal O}(p^3 a^2 r_0)] \; \; \; \; .
\label{bd}
\end{equation}
Thus, for $a>>r_0$, the effects of $V_2$ becomes
comparable to the the effects $V_0$ when $p \sim (a r_0)^{-1/2}$.
This is a signature of the breakdown of the power counting argument.
If $a \rightarrow \infty$,  the momentum scale at which
the power counting breaks down goes to zero.
 
This problem can be avoided quite simply if the $\delta$ function
interactions are regulated from the beginning.  The basic
strategy is to exploit the freedom in choosing $\mu$.  If one begins
with regulated $\delta$ functions, the strength of both $C_0$ and
$C_2$  depend on both the renormalization conditions (fixed by
$a$ and $r_0$) and the regulator mass $\mu$.  In principle, all
physical results should be independent of $\mu$ since it is an
artificial parameter introduced only for convenience.  However, the
full theory is not being solved; within a given approximation
scheme results do depend on $\mu$, albeit only  weakly.  One can exploit
the freedom in choosing $\mu$ to improve the convergence of the
approximation scheme.  An optimal choice of $\mu$ is one which
minimizes the errors associated with truncating the expansion. Thus,
for example, in perturbative QCD treatments of deep inelastic
scattering one chooses the factorization scale $\mu$ to be of order
$Q^2$ in order to avoid large logarithms in the higher order
corrections.  In an analogous fashion, for the present problem one can
fix $\mu$ so as to minimize the higher order corrections of the EFT
expansion.  In particular, one can chose $\mu$ so that $C_2 = 0$.
This is possible for any reasonable regulator since one can fix
$r_0$ and $a$ from the range and depth of the regulated $\delta$
function of $V_0$.  With this optimal regulator $T_2 = T_0$ for all
$p$ and the difficulty of the expansion breaking down at low $p$ is
avoided.  More generally, one expects that if a non-optimal regulator
mass, $\mu$, comparable to or less than $1/r_0$ were chosen then 
$T_2 = T_0 [1 +{\cal O}(p/\mu)]$.  This will be studied in a subsequent
 publication.

There is also a problem with this treatment in the limit $a
\rightarrow 0$.  This corresponds to a zero $T$ matrix at zero energy.
This situation can occur in a nontrivial way if the underlying
potential has both attraction and repulsion whose effects cancel at
zero energy; it can also occur in an attractive potential with a
sufficiently deeply bound state. In general, for scattering problems with
nonzero potentials and $a=0$, the $T$ matrix is zero only for zero
energy.  For a generic interaction tuned to give $a=0$, normal power
counting would lead one to expect that $T \sim p^2/\Lambda^2$. In
contrast, consider Eq.~(\ref{T2}).  As $a$ goes to zero, $T_2$ goes to
zero for all $p$ violating the conventional power counting arguments.
Again this represents a serious difficulty since nothing in the
conventional power counting depends in an obvious way on $a$ being
nonzero.  This problem is also an artifact of imposing an infinite
cutoff.

The author gratefully acknowledges discussions with Daniel Phillips, Manoj
Banerjee and
Ubirajara van Kolck.  This work was supported in part by the U.S.
Department of Energy through grant no. DE-FG02-93ER-40762.

\end{document}